\documentclass[spanish,aps,ams,showpacs]{revtex4}

\usepackage{amsmath}
\usepackage{amssymb}
\usepackage{graphicx}
\usepackage{color}
\usepackage[draft=false]{hyperref}
\usepackage[latin1]{inputenc}

\begin{document}

\title{Charged Annular Disks and Reissner-Nordstr\"{o}m Type\\ Black Holes from Extremal Dust}
\author{F. D. Lora-Clavijo}\email[Email: ]{fadulora@ifm.umich.mx}
\affiliation{Instituto de F\'isica y Matem\'aticas, Universidad Michoacana de San Nicol\'as de Hidalgo, \\
Edificio C-3, Cd. Universitaria, A. P. 2-82, 58040 Morelia,  Michoac\'an, M\'exico
}

\author{P. A. Ospina-Henao}\email[Email: ]{paoloandres14@mail.ustabuca.edu.co}
\affiliation{Universidad Santo Tom\'as, Carrera 18 No. 9 - 27 PBX 6 800 801,\\
 Bucaramanga, Colombia
}

\author{J. F. Pedraza}\email[Email: ]{juan.pedraza@nucleares.unam.mx}
\affiliation{Instituto de Ciencias Nucleares, Universidad Nacional Aut\'onoma de M\'exico, \\
Apartado Postal 70-543, M\'exico Distrito Federal 04510, M\'exico
}
\affiliation{Department of Physics, University of Texas, 1 University Station C1600, \\
Austin, Texas 78712-0264, USA
}

\date{\today}

\begin{abstract}

We present the first analytical superposition of a charged black hole with an annular disk of extremal dust. In order to obtain the solutions, we first solve the Einstein-Maxwell field equations for sources that represent disk-like configurations of matter in confomastatic spacetimes by assuming a functional dependence among the metric function, the electric potential and an auxiliary function, which is taken as a solution of the Laplace equation. We then employ the $Lord$ $Kelvin$ $Inversion$ $Method$ applied to models of finite extension in order to obtain annular
disks. The structures obtained extend to infinity, but their total masses
are finite and all the energy conditions are satisfied. Finally, we
observe that the extremal Reissner-Nordstr\"{o}m black hole can be embedded
into the center of the disks by adding a boundary term in the inversion.

\end{abstract}

\maketitle

\section{Introduction}

Exact solutions of Einstein equations with axial symmetry play an important role in astrophysical applications of general relativity. In particular, disk-like configurations of matter are of great interest, since these structures are present in several systems, such as galaxies, nebulae and accretion disks around compact objects. In fact, there are still important unanswered questions in this field, namely the formation of Active Galactic Nuclei (AGN), X-ray transients and gamma-ray bursts, which are commonly associated with the accretion of matter onto black holes.

It is commonly thought that these sources are able to form collimated,
ultrarelativistic flows (relativistic jets). The exact mechanisms to explain the production of jets are still uncertain,
but they probably involve the interaction between a rotating black
hole, the accretion disk and electromagnetic fields in strong gravitational
fields (see, for example, \cite{Kundt,Krolik,Blandford} and references therein). Thus, an accurate general relativistic model of an AGN would require an exact solution of Einstein-Maxwell field equations that describes a nonlinear
superposition of a Kerr black hole with a stationary disk coupled to electromagnetic
fields. Not even an exact solution of a stationary black hole-disk system has
been found yet.

Hence, the study of systems composed by a thin disk surrounding a central black hole is of great relevance in astrophysics and it is considered an open problem in general relativity. A lot of work has been done in the last years in order to obtain exact solutions of Einstein's equations corresponding
to thin disk-like sources with a central black hole as well as to obtain a better understanding
of the different aspects involved in the dynamics of
these systems (see \cite{SEM01,KHS} for thoroughly reviews on the subject).

Solutions for static thin disks without radial pressure were first studied by Bonnor and Sackfiel \cite{BS} and Morgan and Morgan \cite{MM1} and with radial pressure by Morgan and Morgan \cite{MM2}. Other classes of static thin disks were obtained by \cite{LP1,LEM,BLK,BLP,GL1,GE} and stationary thin disks were studied in  \cite{LP2,BL,PL,GL2}. Models of thin disks with halos made of perfect fluid were considered in \cite{VL1}. The generalization of the $displace$, $cut$ $and$ $reflect$ method for constructing static thick disks was considered by \cite{GL}. Thin disks have been discussed  as sources for Kerr-Newman  field \cite{LBZ,GG1}, magnetostatic axisymmetric fields \cite{LET1} and conformastatic and conformastationary metrics \cite{VL2,KBL}, while models of electrovacuum static counterrotating dust disks were presented in \cite{GG2}. Charged perfect fluid disks were also studied in \cite{VL3}, and  charged perfect fluid disks as sources of  static and  Taub-NUT-type spacetimes in \cite{GG3,GG4}.

However, in the context of a superposition with a black hole, if we consider disks that extend up to
the event horizon, the matter located near the black hole
will have superluminal velocities, as was shown
by Lemos and Letelier \cite{LL1,LL2,LL3}. So, in order to prevent
the appearance of tachyonic matter, the thin disks
must have a central hole with a radius larger than the
photonic radius of the black hole. Then, the boundary
value problem is mathematically
more complicated and thus only very few exact solutions
have been obtained.

This kind of solutions were first
studied by Lemos and Letelier \cite{LL2} for static
axisymmetric spacetimes to obtain exact disk solutions around a Schwarzschild-type
black hole. Since the static axisymmetric Einstein equations are equivalent to the
Laplace equation and a nonlinear quadrature, this supersposition of solutions is
possible there. The main properties of this kind of superposition were extensively
analyzed in a series of papers by Semer\'ak, \u{Z}\'a\u{c}ek and Zellerin \cite{SZZ1,SZZ2,SZ1,SZ2,SEM2,ZS1,SEM3}, by using numerical
computation when needed. Besides these models, only
two other solutions for static thin disks with an inner
edge have been obtained, a first one with inverted
isochrone disks \cite{K1} and a second one for disks with
a power-law density \cite{SEM4}. Also, Zellerin and Semerak \cite{ZS2}, by
using the Belinskii-Zakahrov inverse-scattering method, found a stationary
metric that reduces to the superposition of a disk and a black hole in the
static limit and thus may represent a stationary disk-black hole system, but the analysis of their solution is complicated by the fact that the metric functions cannot be analytically computed. Furthermore, this solution involves an unphysical
supporting surface between the black-hole horizon
and the disk \cite{SEM5}. Finally, general class of stationary
solutions was presented by Klein \cite{K2,K22}, using
the Riemann-surface techniques, in which physically
acceptable black hole disk systems can be found. At this point it is worth to mention that a common feature of all the above mentioned solutions
is that their metric functions can not be fully analytically
computed and most of them present singularities
at the inner edge of the disk. Thus, the analysis of their physical
and mathematical properties is very complicated. Recently, the first fully integrated exact solution of the Einstein equations corresponding
to the superposition of an annular dust disk with a neutral black hole was founded in \cite{memog,memog2}, overcoming most of the problems mentioned above.

In this paper we present the first superposition of a charged black hole with an annular disk made of extremal dust. Although one may intuitively expect that astrophysical objects are neutral, the possibility that compact objects could actually contain a non-vanishing net charge was first pointed out in \cite{R,BH}, where the authors modeled a star as a gas of positive ions and electrons and concluded that, due to their greater kinetic energy, the electrons tend to escape from the star more often than the ions. The star will then acquire a net positive charge. The process will be carried on until the electric field induced in the star stops more electrons from escaping. Recently other mechanisms to induce electric charge into compact objects, in particular into black holes, have been proposed (see for example \cite{DARIO}). In order to obtain the superposition, we first construct two infinite families of solutions of the Einstein-Maxwell equations in conformastatic spacetimes that represent disk-like configurations of matter with a central hole and charge density equal to the mass density, in such a way that the electric repulsions and gravitational attractions are in exact balance. This kind of equilibrium configuration of matter, has been called by some authors $Electrically$ $Counterpoised$ $Dust$ (ECD) and has been studied with some detail, both in classical and relativistic theories \cite{das,BW,B,BB,BB2,GUR,ivanov,VAR}. Also, disk sources for conformastationary metrics have been considered in \cite{KBL}. The method that will be used to construct the corresponding inverse structures is the so-called $Lord$ $Kelvin$ $Inversion$ $Method$, applied to models of finite extension. The structures obtained are extended to infinite, but one can put a cutoff due to the fast decay rate of the densities and their masses are finite. Moreover, the solutions obtained here agrees with all the energy conditions. We observe that the extremal Reissner-Nordstr\"{o}m black hole can be embedded in the kind of spacetimes under consideration and, in particular, we note that such geometry arises naturally in the context of a (Kelvin) inversion as a result of a boundary term. Therefore, the superposition of a disk with the black hole turns out to be straightforward. This superposition can be used as a first approximation of an AGN and, due to its relative simplicity, may be relevant to construct more realistic models.

The paper is organized as follows. In Sec. \ref{sec:ecs}, we present the solutions of the Eisntein-Maxwell equations obtained in \cite{paolo}, for comformastatic spacetimes. Next, in Sec. \ref{sec:emt}, we obtain the surface energy momentum tensor and the surface current density of the relativistic thin disks. In Section \ref{sec:kel}, we show the $Lord$ $Kelvin$ $Method$ and we apply it to the solution obtained in the Sec. \ref{sec:emt}.  We verify indeed, that the solution reduces to the Minkowski one at infinity and we check that this is in agreement with the energy conditions. We present in Sec. \ref{sec:kal} two particular families of solutions that represent sources of charged matter with a central hole. The distributions, which we applied this method to, are the Morgan and Morgan disks (MM) \cite{MM1} and the flat rings introduced by \cite{let2}. Then, in Sec. \ref{sec:RNBH} we show the way to superpose this kind of solutions with an extremal Reissner-Nordstr\"{o}m black hole and in Sec. \ref{exsol} we obtain the first one explicitly. Finally, in Sec \ref{sec:conc}, we present the conclusions of our main results.

\section{\label{sec:ecs}Einstein-Maxwell equations for conformastatic spacetimes}

For conformastatic spacetimes the line element  can be written in  cylindrical coordinates $x^\mu = (t, \varphi, r, z)$ as \cite{SYN}
\begin{equation}
ds^2 = - \ e^{2\lambda} dt^2  +  e^{- 2\lambda} (r^2d\varphi^2+dr^2+dz^2),
\label{eq:metCC}
\end{equation}
where the metric function $\lambda$ depends on $(r,z)$.

The electrovacuum Einstein-Maxwell system of equations, in geo\-me\-trized units such that $c = G
= \mu _{0} = \epsilon _{0} =  1$, is given by
\begin{eqnarray}
&& G_{\mu\nu} = 8 \pi \ T_{\mu\nu},  \label{eq:emep1} \\
&&    \nonumber    \\
&& T_{\mu\nu} = \frac{1}{4 \pi} \left[ F_{\mu \alpha} F_\nu^{\alpha} - \frac{1}{4} g_{\mu\nu} F_{\alpha\beta}
F^{\alpha \beta} \right], \label{eq:emtensor}  \\
&& \nonumber \\
&& {F^{\mu \nu}}_{;\nu}=0, \label{eq:emep2}\\
&& \nonumber \\
&& F_{\mu\nu} =  A_{\nu,\mu} - A_{\mu,\nu}, \label{eq:fab}
\end{eqnarray}
where $A_\mu$ is the electromagnetic four
potential given by  $A_\mu = (- \phi, 0, 0, 0)$. Now, it is  assumed that the electric potential $\phi$ depends also on
$(r,z)$, in such a way that the electrovacuum Einstein-Maxwell system of equations reduces to
\begin{eqnarray}
\nabla^2 \lambda &=& e^{- 2 \lambda} \nabla \phi \cdot \nabla \phi,
\label{eq:eme2} \\
&  & \nonumber \\
\nabla^2 \phi &=& 2 \ \nabla \lambda \cdot \nabla \phi, \label{eq:eme3}\\
 &  & \nonumber \\
\lambda_{,i} \lambda_{,j} &=& e^{- 2 \lambda} \phi_{,i} \ \phi_{,j},
\label{eq:eme1}
\end{eqnarray}
where $i,j = 1,2,3$, and $\nabla$ is the usual differential operator in
cylindrical coordinates.

A particular subclass of solutions for this system of equations was obtained in \cite{paolo},
where the authors considered a functional dependence between the metric function $\lambda$ and the electric potential $\phi$  with respect to an auxiliary function $U$. This auxiliary function is a solution of the Laplace equation. The explicit form of  $\lambda$ and $\phi$ are given by
\begin{equation}
e^{\lambda} = \frac{k_1}{U + k_2}, \label{lambdau}
\end{equation}
\begin{equation}
\phi = \pm \left[e^{\lambda}  - 1 \right], \label{eq:phiu}
\end{equation}
where $k_1$ and $k_2$ are integration constants, which are chosen in such a way that the metric (\ref{eq:metCC}) reduces to Minkowski at infinity. As we can see when $e^{\lambda}=1$ the electric potential is equal to zero.

\section{\label{sec:emt}Energy-momentum tensor and current density}

As we can see from (\ref{lambdau}) and (\ref{eq:phiu}), the solutions of the Einstein-Maxwell equations system (\ref{eq:eme2})-(\ref{eq:eme3}) are expressed in terms of $U$, a solution of the Laplace equation. We shall deal with solutions of the Laplace equation corresponding to a Newtonian surface mass distribution so that, in general, $U$ is a continuos function everywhere but its first $z$-derivative must be discontinuous at $z=0$. Accordingly, in order to obtain the energy-momentun tensor and the current density of the source, we
will express the jump across the disk of the first $z$-derivatives of the metric tensor as
\begin{equation}
 b_{\mu\nu} =  [{g_{\mu\nu,z}}] = 2 {g_{\mu\nu,z}}|_{_{z = 0^+}},
\end{equation}
and the jump across the disk of the electromagnetic field tensor as
\begin{equation}
[F_{z\mu}] = [A_{\mu,z}] = 2 {A_{\mu,z}}|_{_{z = 0^+}},
\end{equation}
where the reflection symmetry of the functions with respect to $z = 0$ has
been used.

Then, by using the distributional approach \cite{PH,LICH,TAUB}, the Einstein-Maxwell equations yield the energy-momentum tensor
\begin{equation}
T^{\mu\nu} = T_+^{\mu\nu} \theta(z) + T_-^{\mu\nu} [1 - \theta(z)] + Q^{\mu\nu} \delta(z),
\label{eq:emtot}
\end{equation}
and a current density
\begin{equation}
J^\mu = I^\mu \delta(z),   \label{eq:courrent}
\end{equation}
where $\theta(z)$ and $\delta (z)$ are respectively the Heaveside and Dirac distributions with support on $z = 0$. Here $T_\pm^{\mu\nu}$ are the electromagnetic energy-momentum tensors as defined by (\ref{eq:emtensor}) for the $z \geq 0$ and $z \leq 0$ regions respectively, whereas
\begin{equation}
16 \pi Q^\mu_\nu = b^{\mu z}\delta^z_\nu - b^{zz}\delta^\mu_\nu + g^{\mu z}b^z_\nu -
g^{zz}b^\mu_\nu   +  b^\alpha_\alpha (g^{zz}\delta^\mu_\nu - g^{\mu z}\delta^z_\nu)
\end{equation}
gives the part of the energy-momentum tensor corresponding to the disk source, and
\begin{equation}
4 \pi I^\mu  =[F^{\mu z}]
\end{equation}
is the contribution of the disk source to the current density. Now, the ``true'' surface energy-momentum tensor of the disk, $S_{\mu\nu}$, and the
``true'' surface current density, $j^\mu$, can be obtained through the
relationships
\begin{eqnarray}
S_{\mu\nu} &=& \int Q_{\mu\nu} \ \delta (z) \ ds_n \ = \ e^{ - \lambda} \ Q_{\mu\nu} \ ,    \\
&   &           \nonumber      \\
j^\mu &=& \int I^\mu \ \delta (z) \ ds_n \ = \  e^{- \lambda} I^\mu ,
\end{eqnarray}
where $ds_n = \sqrt{g_{zz}} \ dz$ is the proper length in
the  normal direction to the disk.

Regarding the metric in Eq.(\ref{eq:metCC}), the only non-zero component of $Q^\mu_\nu$ is
\begin{equation}
Q^0_0 = - \frac{e^{2\lambda} \lambda_{,z}}{2 \pi}, \label{eq:emt}
\end{equation}
whereas the only non-zero component of $I^\mu$ is
\begin{equation}
I^0 = - \frac{\phi_{,z}}{2 \pi}.
\end{equation}
Thus, the only non-zero component of the surface energy-momentum tensor $S^\mu_\nu$ is
\begin{equation}
S^0_0 = - \frac{e^{\lambda} \lambda_{,z}}{2 \pi}, \label{eq:emts}
\end{equation}
and the only non-zero component of the surface current density $j^\mu$ is
\begin{equation}
j^0 = \ - \frac{e^{- \lambda} \phi_{,z}}{2 \pi},  \label{eq:37}
\end{equation}
where all the quantities are evaluated at $z = 0^+$.

The surface energy-momentum tensor and the surface current density
of the disk, measured in rest frame of the dust, can be written as
\begin{eqnarray}
S^{\mu\nu} &=& \epsilon V^\mu V^\nu , \\
 &	&	\nonumber	\\
j^\mu &=& \sigma V^\mu ,
\end{eqnarray}
where
\begin{equation}
V^\mu = e^{-\lambda} (1, 0, 0, 0 ),
\end{equation}
is the velocity vector of the matter distribution. Thus, the energy
density and the charge density of the distribution of matter are given by
\begin{eqnarray}
\epsilon &=&  \frac{e^\lambda \lambda_{,z}}{2 \pi} , \label{eq:epsi} \\
 & & 	\nonumber	\\
\sigma &=&  - \frac{\phi_{,z}}{2 \pi} \label{eq:sigma} ,
\end{eqnarray}
respectively. Now, by using Eq.(\ref{eq:phiu}), the expression
(\ref{eq:sigma}) can be written as
\begin{equation}
\sigma = \mp \ \epsilon  \label{eq:sigma*},
\end{equation}
that, when static charged dust matter is included, the above conditions imply that the charge density of the disks must be equal to the mass density, with the possible exception of a sign. This kind of Einstein-Maxwell-extremal-dust solutions are
generically called Majumdar-Papapetrou solutions \cite{MAD5,PAP5} and present the interesting feature that they permit the construction of configurations of matter in static equilibrium.

In particular, electrovacuum solutions with
$\sigma=\epsilon=0$, in some cases, reduce to a configuration of many extremal
Reissner-Nordstr\"om  black holes, as was fully explored by Hartle and Hawking \cite{hartlehawking}. Solutions with surface mass density have been studied by \cite{paolo}, and
solutions with volumetric mass density have been examined in the papers of Das~\cite{das},
Bonnor~\cite{B,BB,BB2,BW}, and others (see Ref.~\cite{GUR,ivanov,VAR} and
the references cited therein).

Now, in the context of classical general relativity, the
energy-momentum tensor is suppossed to fulfill certain requirements, which are embodied in
the {\it weak, strong and dominat energy conditions} \cite{HE}. Indeed, for the
case of a dust source, all these conditions reduce to the single condition that the
energy density be greater or equal to zero, $\epsilon \geq 0$. On the other
hand, from (\ref{lambdau}) and (\ref{eq:epsi}), we have that the energy density can be
written as
\begin{equation}
\epsilon = - \frac{k_1 \Sigma}{(U + k_2)^2} , \label{eq:epsig}
\end{equation}
where
\begin{equation}
\Sigma = \frac{U_{,z}}{2 \pi},
\end{equation}
is the Newtonian mass density of a disk-like source whose gravitational potential
is given by $U$. Accordingly, if the Newtonian mass density $\Sigma$ is
non-negative everywhere, the corresponding relativistic energy density $\epsilon$
will be non-negative everywhere only if we take $k_1 < 0$. Therefore, in order that
the energy-momentum tensor of the disks agrees with all the energy
conditions, the $C_{2n}$ constants in (\ref{eq:gensol}) must be properly chosen
in such a way that $\Sigma \geq 0$. Furthermore, if we have a Newtonian
potential $U$ that is negative everywhere, as is expected for a compact
Newtonian source, the energy density $\epsilon$ of the disk will be non-singular
everywhere.

\section{\label{sec:kel}Axisymmetric thin disk solutions with a central hole}

One of the most relevant properties of the energy density (\ref{eq:epsig}) is that it depends on $r$ only through the potential-density pair corresponding to a Newtonian thin disk. So, in order to obtain
solutions that correspond to relativistic thin disks with a central hole we must use solutions of the
Laplace equation that properly describe Newtonian sources with this behavior.

A simple way to construct this kind of solutions is by using the Kelvin inversion method with finite disk-like systems. In the literature we can find several models representing such systems, see for example \cite{KAL,MEST,MILG,MILG2,GR,PRG,EAR}. Thin disks of finite extension can be obtained by solving the Laplace equation in oblate spheroidal coordinates $(\xi,\eta,\varphi)$. These coordinates are related with the usual cylindrical coordinates by the relations
\begin{eqnarray}
r^{2} &=& a^{2}(1 + \xi^{2})(1 - \eta^{2}), \label{eq:ciloblatas1} \\
& & 	\nonumber \\
z &=& a \xi \eta, \label{eq:ciloblatas2}
\end{eqnarray}
with the  $0 \leq \xi < \infty$ and $-1 \leq \eta < 1$. The general solution for a disk of radius $a$ with axial symmetry is given by \cite{BAT}
\begin{equation}
U(\xi,\eta) = - \sum_{n=0}^{\infty} C_{2n} q_{2n}(\xi) P_{2n}(\eta),
\label{eq:gensol}
\end{equation}
where $P_{2n}(\eta)$ and $q_{2n}(\xi)=i^{2n+1}Q_{2n}(i\xi)$ are the
usual Legendre polynomials and the Legendre functions of the second
kind respectively, and $C_{2n}$ are arbitrary constants. With this
general solution for the gravitational potential, the surface
mass density takes the form
\begin{equation}
\Sigma(r) = \frac{1}{2 \pi a \eta}\sum_{n=0}^{\infty} C_{2n}(2n+1) q_{2n+1}(0) P_{2n}(\eta).
\label{eq:gensols}
\end{equation}

Now, we use the Kelvin inversion theorem that in cylindrical coordinates
states that if the potential-density pair $U(r,z)$, $\rho(r,z)=\Sigma(r)\delta(z)$ is a solution of Poisson equation, the pair
\begin{equation}
\tilde{U}(r,z)=\frac{a}{\sqrt{r^2+z^2}}U\left(\frac{a^2 r}{r^2+z^2},\frac{a^2 z}{r^2+z^2}\right),\label{eq:kel1}
\end{equation}
\begin{equation}
\tilde{\rho}(r,z)=\tilde{\Sigma}(r)\delta(z),
\end{equation}
where
\begin{equation}
\tilde{\Sigma}(r)=\left(\frac{a}{r}\right)^3 \Sigma\left(\frac{a^2}{r}\right),
\end{equation}
is also a solution of the same equation. Hence, under these transformations, the energy density given by (\ref{eq:epsig}) becomes
\begin{equation}
\tilde{\epsilon}(r)=- \frac{k_1 \tilde{\Sigma}}{(\tilde{U} + k_2)^2}=-\frac{a^3 k_1 \Sigma(a^2/r)}{ r^3 \left(\frac{a}{r}U\left(\frac{a^2}{r},0\right) + k_2\right)^2}, \label{eq:newden}
\end{equation}
whereas the metric function $\lambda$ transforms as
\begin{equation}
e^{\tilde{\lambda}}(r,z) = \frac{k_1}{\tilde{U} + k_2}=\frac{k_1}{\frac{a}{\sqrt{r^2+z^2}}U\left(\frac{a^2 r}{r^2+z^2},\frac{a^2 z}{r^2+z^2}\right) + k_2}. \label{newlam}
\end{equation}

On the other hand, in order to have an appropriated behavior at infinity, we must impose an additional condition to $k_1$ and $k_2$ in such a way that we obtain an asymptotically flat spacetime. So, we will require that $e^{\tilde{\lambda}} = 1$ at infinity or
\begin{equation}
k_1=k_2,\label{eq:39}
\end{equation}
given that $\tilde{U}$ vanishes there, as we can see in Eq.(\ref{eq:kel1}). Moreover, under the condition in Eq.(\ref{eq:39}), the electric potential given by (\ref{eq:phiu}) also vanishes at infinity.

Here, if we chose solutions to the laplace equation corresponding to finite distributions of matter without singularities, then the Newtonian potential $U$ is negative everywhere and the corresponding energy density will be nonsingular if $k_2<0$. With this condition, $k_1<0$ so if the Newtonian mass distribution is always positive, it is easy to see that the relativistic energy density of the disk will be also positive. However, note that the factor $a/\sqrt{r^2+z^2}$ in (\ref{eq:kel1}) introduces singularity at the origin unless that $U=0$ at infinity. One can easily check that Eq.(\ref{eq:gensol}) implies that $U$ vanishes at $\eta=0$ and $\xi\rightarrow\infty$, but in principle it is possible to add a constant to change its asymptotic behavior. We will assume in first place that this boundary condition is satisfied, but we shall show later that it is precisely this term that will give rise to an extremal Reissner-Nordstr\"{o}m black hole at the center of the disk.

Now, it is possible to show that, although the new models have infinite extension, the total masses remain finite. As the disks are made of dust we have that $p=0$ and the mass density is equal to the energy density. The total mass can be computed from the Komar integral which, in the static axisymmetric case (\ref{eq:metCC}), takes the following form
\begin{equation}
\mathcal{M}= 2\pi\int_a^{\infty}\tilde{\epsilon}(r) e^{- \tilde{\lambda}}r dr,
\end{equation}
Now, by taking into account equations (\ref{eq:newden}), (\ref{newlam}) and (\ref{eq:39}) we find that
\begin{equation}
\mathcal{M}=- 2\pi \int_a^{\infty}\frac{\tilde{\Sigma}(r)}{\tilde{U}(r) + k1}r dr=-2\pi \int_a^{\infty}\frac{\left(\frac{a}{r}\right)^3 \Sigma\left(\frac{a^2}{r}\right)}{\left ( \frac{a}{r}U(\frac{a^2}{r}) + k_1 \right)}rdr.
\end{equation}
Now, we change the variable of integration $r=a^2/R$, and  write the mass of the disk as
\begin{equation}
\mathcal{M}=-2\pi a\int_0^{a}\frac{\Sigma(R)}{\frac{R}{a}U(R) + k_1}dR. \label{masa}
\end{equation}
Hence, if we have a well-behaved Newtonian surface mass density and potential, $\epsilon$ given by (\ref{eq:epsig}) will be non-singular under the condition (\ref{eq:39}) and, the total mass in Eq.(\ref{masa}) will be finite.

In conclusion, we obtain that the general expression for the energy density, which is in complete agreement with all the energy conditions, can be expressed as
\begin{equation}
\tilde{\epsilon}(r)=-\frac{k \tilde{\Sigma}}{(\tilde{U} + k)^2},\label{eq:epsigt}
\end{equation}
where we have set $k_1=k_2=k$ and with the conditions $\Sigma\geq0$ and $k<0$.

\subsection{\label{sec:kal}Particular Solutions}

Now we shall restrict our general model by considering particular
families of solutions by specifying the constants $C_{2n}$ in such a way that the corresponding energy densities have the characteristics we sketched previously. At first instance we shall deal with the MM family, which was introduced in \cite{MM1} (see also \cite{GR}) and whose first member is the well-known Kalnajs disk \cite{KAL}. This family are characterized by its well-behaved surface mass density, labeled with the positive integer $m\geq1$, and given by
\begin{equation}
\Sigma^{(m)}(r)=\frac{(2m+1)M}{2\pi a^2}\left(1-\frac{r^2}{a^2}\right)^{m-\frac{1}{2}},\label{denskal}
\end{equation}
where $M$ is the Newtonian mass of the disk and $a$ is the radius. For each member of the family, the constants $C_{2n}$ are given by
\begin{equation}
C_{2n}=\frac{M \pi^{1/2}(4n+1)(2m+1)!}{a 2^{2m+1}(2n+1)(m-n)!\Gamma(m+n+\frac{3}{2})q_{2n+1}(0)},
\end{equation}
for $n\leq m$, and $C_{2n}=0$ for $n>m$. So, by using this constants in the general solution
(\ref{eq:gensol}) it is easy to see that the gravitational potential of each disk $U^{(m)}$ will be negative everywhere, as was mentioned in the previous section. The closed expressions corresponding to the first
four members are given by
\begin{subequations}\begin{align}
U^{(1)}(\xi,\eta) &= - \frac{M}{a} [ \cot^{-1}\xi  + A (3\eta^{2} - 1)],
\label{eq:4.22}   \\
U^{(2)}(\xi,\eta) &= - \frac{M}{a} [ \cot^{-1} \xi + \frac{10 A}{7}
(3\eta^{2} - 1) + B ( 35 \eta^{4} - 30 \eta^{2} + 3)], \label{eq:4.23}  \\
U^{(3)}(\xi,\eta) &= - \frac{M}{a} [ \cot^{-1} \xi + \frac{5 A}{3} (3
\eta^{2} - 1) + \frac{9 B}{11} (35 \eta^{4} - 30 \eta^{2} + 3)
\nonumber   \\
& \quad \quad  + C (231 \eta^{6} - 315 \eta^{4} + 105 \eta^{2} - 5) ], \\
U^{(4)}(\xi,\eta) &= - \frac{M}{a} [ \cot^{-1} \xi + \frac{20 A}{11} (3
\eta^{2} - 1)  +  \frac{162 B}{143} (35 \eta^{4} - 30 \eta^{2} + 3)
\nonumber   \\
& \quad \quad  + \frac{4C}{11} (231 \eta^{6} - 315 \eta^{4} + 105 \eta^{2} -
5) + D(6435 \eta ^8-12012 \eta ^6\nonumber \\
&\quad \quad +6930 \eta ^4-1260 \eta ^2+35)],
\end{align}\end{subequations}
with
\begin{subequations}\begin{align}
A &= \frac{1}{4} [(3\xi^{2} + 1) \cot^{-1} \xi - 3 \xi ], \\
B &= \frac{3}{448} [ (35 \xi^{4} + 30 \xi^{2} + 3) \cot^{-1} \xi - 35 \xi^{3} -
\frac{55}{3} \xi ], \\
C &= \frac{5}{8448} [ (231 \xi^{6} + 315 \xi^{4} + 105 \xi^{2} + 5) \cot^{-1}
\xi - 231 \xi^{5} - 238 \xi^{3} - \frac{231}{5} \xi ], \\
D &= \frac{7}{2342912} [(6435 \xi ^8+12012 \xi ^6+6930 \xi
   ^4+1260 \xi ^2+35) \cot ^{-1}\xi\nonumber \\
   &\quad \quad -6435 \xi ^7-9867 \xi
   ^5-4213 \xi^3-\frac{15159  }{35}\xi].
\end{align}\end{subequations}
It is possible to check that
\begin{equation}
\lim_{\xi\rightarrow\infty}U^{(m)}(\xi,0)=0.
\end{equation}
We restrict our attention to these four members but the behavior of the remaining models ($m\geq 5$) can be inferred from the features
characterizing $m=1,2,3,4$.

With the above values for the constants $C_{2n}$, we can easily compute the corresponding energy density $\tilde{\epsilon}^{(m)}$ of the inverted disks by using equation (\ref{eq:epsigt}). Now, in order to show graphically the behavior of each particular model, we first introduce dimensionless quantities through the relations
\begin{eqnarray}
\hat{U}^{(m)}(\hat{r}) &=& \frac{a \tilde{U}^{(m)} ({\hat r})}{M}, \\
	&&	\nonumber	\\
{\hat \Sigma}^{(m)} ({\hat r}) &=& \frac{ a^2 \tilde{\Sigma}^{(m)} ({\hat r})}{M}, \\
	&&	\nonumber	\\
{\hat \epsilon}^{(m)} ({\hat r}) &=&  a \tilde{\epsilon}^{(m)} ({\hat r}),
\end{eqnarray}
where ${\hat r} = r/a$, and $\tilde{U}^{(m)} ({\hat r})$ is
evaluated at $z = 0^+$. Accordingly, the dimensionless energy density ${\hat
\epsilon} ({\hat r})$ can be written as
\begin{equation}
{\hat \epsilon}^{(m)} ({\hat r}) = - \frac{{\hat k}{\hat \Sigma}^{(m)} ({\hat
r})}{[{\hat U}^{(m)} ({\hat r}) + {\hat k}]^2},\label{sindim}
\end{equation}
with ${\hat k} = a k/M$.

Then, by using the above expressions and the values
of the $C_{2n}$ constants corresponding to the MM disks, we obtain the following
expressions for the first four members of the family

\begin{eqnarray}
{\hat \epsilon}^{(1)} &=& -\frac{3\hat{k} \left(1-\frac{1}{\hat{r}^{2}}\right)^{1/2}}{2 \pi  \hat{r}^3 \left[\frac{3 \pi  }{8 \hat{r}^3}\left(2\hat{r}^2-1\right)-\hat{k}\right]^2 }, \\
	&&	\nonumber	\\
{\hat \epsilon}^{(2)} &=& -\frac{5\hat{k} \left(1-\frac{1}{\hat{r}^2}\right)^{3/2}}{2 \pi  \hat{r}^3 \left[\frac{15 \pi }{128 \hat{r}^5} \left(8 \hat{r}^4-8 \hat{r}^2+3\right)-\hat{k}\right]^2}, \\
	&&	\nonumber	\\
{\hat \epsilon}^{(3)} &=& -\frac{7\hat{k} \left(1-\frac{1}{\hat{r}^2}\right)^{5/2}}{2 \pi  \hat{r}^3 \left[\frac{35 \pi  }{512 \hat{r}^7}\left(16 \hat{r}^6-24 \hat{r}^4 + 18  \hat{r}^2 - 5\right)-\hat{k}\right]^2}, \\
	&&	\nonumber	\\
{\hat \epsilon}^{(4)} &=& -\frac{9\hat{k} \left(1-\frac{1}{\hat{r}^2}\right)^{7/2}}{2 \pi  \hat{r}^3 \left[\frac{315  \pi  }{32768  \hat{r}^9}\left(128 \hat{r}^8-256 \hat{r}^6+288 \hat{r}^4-160 \hat{r}^2+35\right)-\hat{k}\right]^2}.
\end{eqnarray}

\begin{figure*}[!ht]
\begin{center}
\includegraphics[width=4.8in]{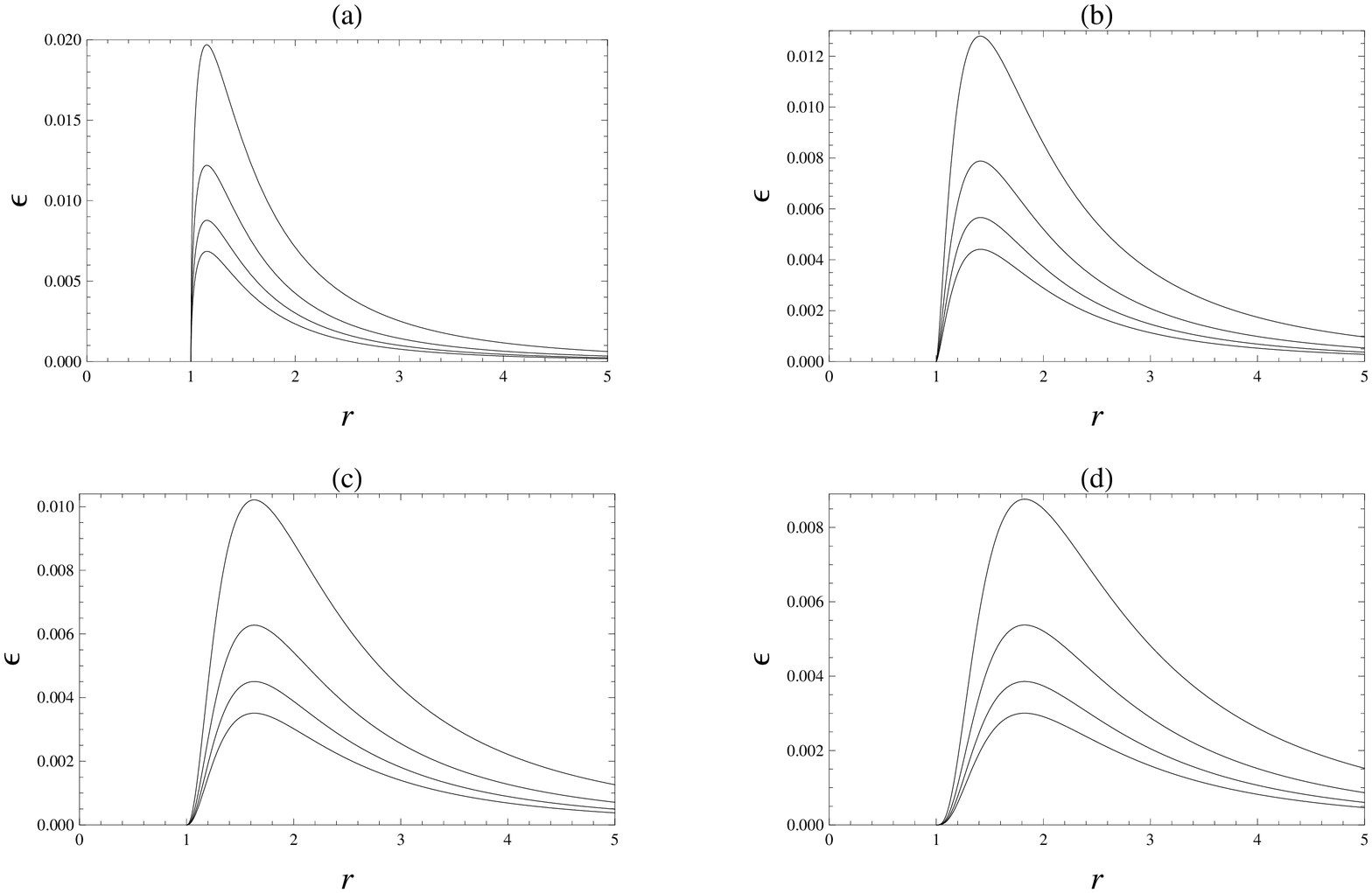}
\caption{Dimensionless surface energy density ${\hat
\epsilon}^{(m)}$ as a function of ${\hat r}$ for the first four members of the family with (a) $m=1$; (b) $m=2$; (c) $m=3$; (d) $m=4$. In each case, we plot ${\hat \epsilon}^{(m)} ({\hat r})$ for $0 \leq {\hat r} \leq 5$ with different values for the parameter ${\hat
k}$. The uppest curve of each plot corresponds to ${\hat k} = -5$, and
then ${\hat k} = -10$, $-15$ and $-20$ for the lowest curve.}\label{fig:1}
\end{center}
\end{figure*}

\begin{figure*}[!ht]
\begin{center}
\includegraphics[width=7cm]{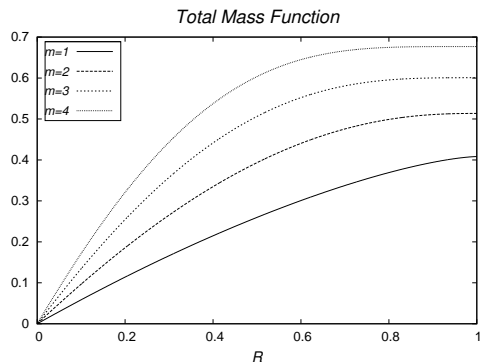}
\caption{In this plot, we present the mass function for the first four disk models of the family with $a=1$, $G=1$, $k=-5$ and $M=1$. As we can see, all the masses converge to a finite value.}\label{fig:masa}
\end{center}
\end{figure*}

\begin{figure}[!ht]
\begin{center}
\includegraphics[width=7cm]{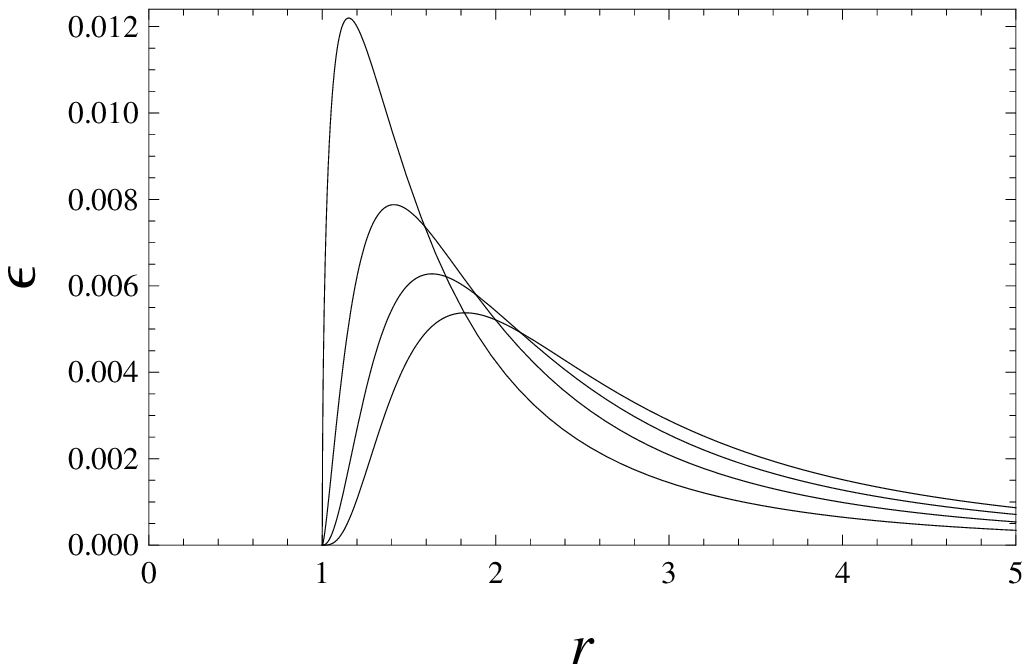}
\caption{Dimensionless surface energy density ${\hat
\epsilon}^{(m)}$ as a function of ${\hat r}$ for the first four disk models of the family with $m
= 1$, $2$, $3$ and $4$. Here we plot ${\hat \epsilon}^{(m)} ({\hat r})$ for $0 \leq {\hat r} \leq 5$ setting ${\hat k}=-10$ for each curve.}\label{fig:2}
\end{center}
\end{figure}

In Fig. \ref{fig:1}, we plot the dimensionless surface energy density $\hat{\epsilon}^{(m)}$ as a function of $\hat{r}$ for the first four members of the family with $m=1, 2, 3$ and $4$. In each case, we plot $\hat{\epsilon}^{(m)}(\hat{r})$ for different values of the parameter $\hat{k}$. The top curve of each plot corresponds to $\hat{k}=-5$ whereas the others correspond to $\hat{k}=-10,-15$ and $-20$ for the lowest curve. As we can see,
in all these cases the energy density is everywhere positive. The curves have a central hole for $\hat{r}\leq1$, then it reaches a maximum and finally it decreases quickly and vanishes at infinity. Here, there are some characteristics of the plots that we want to point out. First, we realize that for each case, the maximum is the highest for the first model ($m=1$) and then, it decreases as we consider the following disks ($m=2,3,...$). Moreover, unlike the finite models studied by \cite{paolo}, the behavior of the energy density with respect to the integration constant $\hat{k}$ is the same for all models: $\hat{\epsilon}$ increases as $|\hat{k}|$ decreases for all values of $\hat{r}$, and the plots corresponding to different values of $\hat{k}$ do not intersect each other. Finally, we can see that the maximum of the first model is the closest to the edge ($\hat{r}=1$) while the other curves present the maximum at a larger distance to the edge as $m$ increases. The energy density of these models decays in a similar way as the Plummer-Kuzmin disks, i.e, as $1/\hat{r}^3$, not too fast. The total masses can be computed inserting (\ref{denskal}) into equation (\ref{masa}). However, it is not possible to solve the integral analytically and hence, we proceeded to do it numerically for each family member. 

In Fig. \ref{fig:masa}, we plot the total mass function for the first four models of the family. As was expected, the total masses are finite in spite of the fact that the models have infinite extension.

On the other hand, we plot in Figure \ref{fig:2} the dimensionless surface energy density for the models $m=1,2,3$ and $4$ keeping $\hat{k}$ at a fixed value. From this graphics it is important to notice that the models with large $m$ decreases more slowly than the first models, but they have a lower maximum. Consequently we see that, in this case, each curve intersects the others at a certain point.\\

Now let's consider disks of the same radius $a$ and decreasing mass
\begin{equation}
M^{(n)}=\frac{2\pi  a^2 \Sigma_c}{2n+1},
\end{equation}
where $\Sigma_c$ is a constant taken equal for all disks of the MM family, $n=1,2,3,...$. For this family of disks we have that the corresponding surface density is
\begin{equation}
\Sigma^{(n)}=\Sigma_c \left(1-\frac{r^2}{a^2}\right)^{\frac{1}{2}}\left(1-\frac{r^2}{a^2}\right)^{n-1}.
\end{equation}
If we consider the superposition given by
\begin{eqnarray}
\Sigma^{(m)}_r &=& \sum_{n=0}^{m}\frac{(-1)^{m-n}m!}{n!(m-n)!}\Sigma^{(m+1-n)}, \\
& & 	\nonumber \\
&=& \Sigma_c\left(1-\frac{r^2}{a^2}\right)^{\frac{1}{2}}\frac{r^{2m}}{a^{2m}},
\end{eqnarray}
we obtain the family of rings recently introduced by \cite{let2}. We have that all these superpositions give disks of radius $a$ with zero density on their centers, i.e., disks with a hole in their centers with a residual density that is smaller for larger $m$. The potentials associated to these flat rings can be found by using a superposition with the same coefficients as the ones used for the densities.

For this family of models, we introduce the following dimensionless quantities
\begin{eqnarray}
\hat{U}^{(m)}(\hat{r}) &=& \frac{\tilde{U}^{(m)} ({\hat r})}{a \Sigma_c}, \\
	&&	\nonumber	\\
{\hat \Sigma}^{(m)} ({\hat r}) &=& \frac{ \tilde{\Sigma}^{(m)} ({\hat r})}{\Sigma_c}, \\
	&&	\nonumber	\\
{\hat \epsilon}^{(m)} ({\hat r}) &=&  a \tilde{\epsilon}^{(m)} ({\hat r}),
\end{eqnarray}
where ${\hat r} = r/a$, and $\tilde{U}^{(m)} ({\hat r})$ is
evaluated at $z = 0^+$. Accordingly, the dimensionless energy density ${\hat
\epsilon} ({\hat r})$ can be written in the same way as equation (\ref{sindim}), but using the new definitions for $\hat{U}$ and ${\hat \Sigma}$.

Then, by using the above expressions for the flat rings previously introduced, we obtain the following
expressions for the first four members of the family

\begin{eqnarray}
{\hat \epsilon}^{(1)} &=& -\frac{\hat{k} \sqrt{1-\frac{1}{\hat{r}^2}}}{\hat{r}^5\left[\frac{\pi ^2}{64\hat{r}^5} \left(8 \hat{r}^4+8 \hat{r}^2-9\right)-\hat{k}\right]^2}, \\
	&&	\nonumber	\\
{\hat \epsilon}^{(2)} &=& -\frac{ \hat{k} \sqrt{1-\frac{1}{\hat{r}^2}} }{\hat{r}^7\left[\frac{\pi ^2}{256 \hat{r}^7} \left(16 \hat{r}^6+8 \hat{r}^4+18 \hat{r}^2-25\right)- \hat{k}\right]^2}, \\
	&&	\nonumber	\\
{\hat \epsilon}^{(3)} &=& -\frac{\hat{k} \sqrt{1-\frac{1}{\hat{r}^2}} }{\hat{r}^9\left[\frac{\pi ^2}{16384 \hat{r}^9} \left(640 \hat{r}^8+256 \hat{r}^6+288 \hat{r}^4+800 \hat{r}^2-1225\right)-\hat{k}\right]^2}, \\
	&&	\nonumber	\\
{\hat \epsilon}^{(4)} &=& -\frac{\hat{k} \sqrt{1-\frac{1}{\hat{r}^2}}}{\hat{r}^{11} \left[\frac{\pi ^2}{65536 \hat{r}^{11}} \left(1792 \hat{r}^{10}+640 \hat{r}^8+576 \hat{r}^6+800 \hat{r}^4+2450 \hat{r}^2-3969\right)-\hat{k}\right]^2}. \nonumber \\
\end{eqnarray}

\begin{figure*}[!ht]
\begin{center}
\includegraphics[width=4.8in]{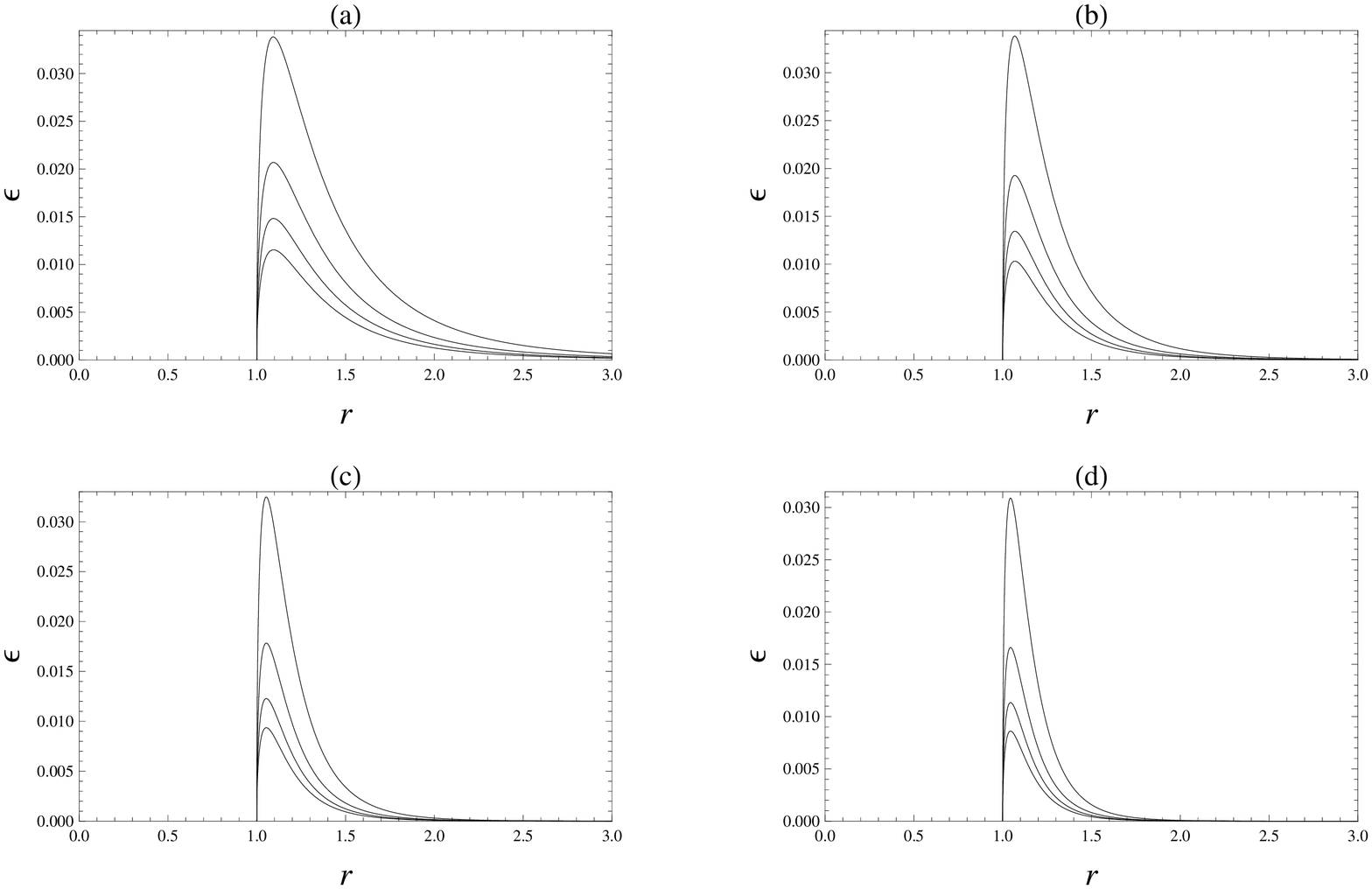}
\caption{\label{figure3}Dimensionless surface energy density ${\hat
\epsilon}^{(m)}$ as a function of ${\hat r}$ for the first four members of the second family with (a) $m=1$; (b) $m=2$; (c) $m=3$; (d) $m=4$. In each case, we plot ${\hat \epsilon}^{(m)} ({\hat r})$ for $0 \leq {\hat r} \leq 3$ with different values for the parameter ${\hat
k}$. The uppest curve of each plot corresponds to ${\hat k} = -5$, and
then ${\hat k} = -10$, $-15$ and $-20$ for the lowest curve.}
\end{center}
\end{figure*}

\begin{figure*}[!ht]
\begin{center}
\includegraphics[width=7cm]{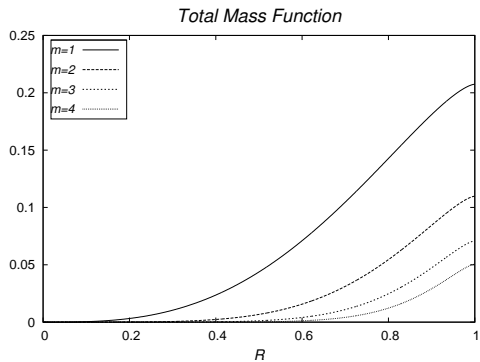}
\caption{In this plot, we present the mass function for the first four disk models of the family with $a=1$, $G=1$, $k=-5$ and $\Sigma_c=1$. As we can see, all the masses are finite.}\label{fig:masa2}
\end{center}
\end{figure*}

\begin{figure}[!ht]
\begin{center}
\includegraphics[width=7cm]{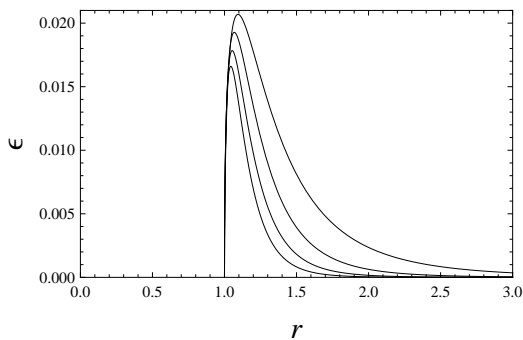}
\caption{\label{figure4}Dimensionless surface energy density ${\hat
\epsilon}^{(m)}$ as a function of ${\hat r}$ for the first four disk models of the second family with $m
= 1$, $2$, $3$ and $4$. Here we plot ${\hat \epsilon}^{(m)} ({\hat r})$ for $0 \leq {\hat r} \leq 3$ setting ${\hat k}=-10$ for each curve.}
\end{center}
\end{figure}

In Fig. \ref{figure3}, we plot the dimensionless surface energy
density $\hat{\epsilon}^{(m)}$ as a function of $\hat{r}$ for the first four members of the family with $m=1, 2, 3$ and $4$. In each case, we plot $\hat{\epsilon}^{(m)}(\hat{r})$ for different values of the parameter $\hat{k}$. The top curve of each plot corresponds to $\hat{k}=-5$ whereas the others correspond to $\hat{k}=-10,-15$ and $-20$ for the lowest curve. The energy density is everywhere positive as was expected. The curves have a central hole for $\hat{r}\leq1$, like the first family, then they reach a maximum and finally they decrease very quickly and vanish at infinity. The energy density of these models decays as $\hat{\epsilon}^{(m)} \sim 1/\hat{r}^{2m+3}$. As before, the total mass functions are computed numerically and in Fig. \ref{fig:masa2}, we show the behavior of the first four models for this family.


The behavior of these plots are very similar to those showed in Fig. \ref{fig:1}. They present the highest maximum for the model with $m=1$ with the maximums decreasing as we consider the following disks. Moreover $\hat{\epsilon}$ increases as $|\hat{k}|$ decreases for all values of $\hat{r}$. Now, due to the fact that the densities decay very fast, we can put a clear cutoff and, thus, consider this family of models as finite flat rings.

On the other hand, we plot in Fig. \ref{figure4} the dimensionless surface energy density for the models $m=1,2,3$ and $4$ keeping $\hat{k}$ at a fixed value. In this case, we see that as $m$ increases the maximum decreases but the variation is slower than in the first family. Moreover, the decay is faster for larger values of $m$ and as a result, each curve does not intersect the others, in contrast with the behavior of Fig. \ref{fig:2}.

\section{\label{sec:RNBH}Solutions with an Extremal Reissner-Nordstr\"{o}m Black Hole}

The well-known metric of the Reissner-Nordstr\"{o}m (RN) black hole spacetime in Schwarzschild coordinates is given by
\begin{eqnarray}
ds^2 = -\left[{(\rho-\rho_+)(\rho-\rho_-) \over \rho^2}\right] dt^2 +\left[{(\rho-\rho_+)(\rho-\rho_-)
    \over \rho^2}\right]^{-1} d\rho^2 + \rho^2 d\Omega^2,
\label{rn}
\end{eqnarray}
with horizons appearing at $\rho=\rho_{\pm}$, where $\rho_{\pm}=M_{\mbox{\scriptsize bh}}\pm\sqrt{M_{\mbox{\scriptsize bh}}^2-Q^2}$. The
line $Q^2 = M_{\mbox{\scriptsize bh}}^2$ in the parameter space of electrovacuum solutions is referred to as the
extremal RN spacetime. For such spacetimes the metric is reduced to
\begin{equation}
ds^2=-\left(1-\frac{M_{\mbox{\scriptsize bh}}}{\rho}\right)^2dt^2+\left(1-\frac{M_{\mbox{\scriptsize bh}}}{\rho}\right)^{-2}d\rho^2
+\rho^2(d\theta^2+\sin\theta^2d\varphi^2). \label{3c}
\end{equation}
Similar to the Schwarzschild spacetime, the RN spacetime is spherically
symmetric and static. The metric (\ref{3c}) can be rewritten in harmonic coordinates as
\begin{equation}
    ds^2=-\left(1+\frac{M_{\mbox{\scriptsize bh}}}{R}\right)^{-2}dt^2+ \left(1+\frac{M_{\mbox{\scriptsize bh}}}{R}\right)^2\left[dR^2+R^2
       (d\theta^2+\sin^2\theta d\varphi^2) \right],
\label{harmonicmetricspherical}
\end{equation}
where $\rho=R+M_{\mbox{\scriptsize bh}}$. Some comment on the range of the radial coordinates is in order here. $R>0$ covers only the region
outside the horizon; the region inside the horizon is obtained by continuing the solution to the range $-M_{\mbox{\scriptsize bh}} \le R <0$.

Now, writing the same metric in cylindrical coordinates, we obtain
\begin{equation}
ds^2=-\left(1+\frac{M_{\mbox{\scriptsize bh}}}{\sqrt{r^2+z^2}}\right)^{-2}dt^2+ \left(1+\frac{M_{\mbox{\scriptsize bh}}}{\sqrt{r^2+z^2}}\right)^2(dr^2+dz^2+r^2d\varphi^2),\label{rncyl}
\end{equation}
which coincides with (\ref{eq:metCC}) under the identification
\begin{equation}
e^\lambda=\left(1+\frac{M_{\mbox{\scriptsize bh}}}{\sqrt{r^2+z^2}}\right)^{-1}.
\end{equation}
This implies that given a potential $U=M_{\mbox{\scriptsize bh}}k/\sqrt{r^2+z^2}$, and using Equation (\ref{lambdau}), one recover the extremal RN spacetime.

Now then, lets suppose that we add a constant to the potential given by (\ref{eq:gensol}) obtaining $U'(r,z)=U(r,z)+C$. In this case $\Sigma'(r)=\Sigma(r)$ with $C$ representing the potential at infinity. By choosing $C=M_{\mbox{\scriptsize bh}}k/a$, the Kelvin inversion theorem tell us that the inverted potential is now
\begin{equation}
\tilde{U}'(r,z)=\tilde{U}(r,z)+\frac{M_{\mbox{\scriptsize bh}}k}{\sqrt{r^2+z^2}},
\end{equation}
whereas the inverted density changes only by a delta factor \cite{comment}
\begin{equation}
\tilde{\Sigma}'(r)=\tilde{\Sigma}(r)+\frac{M_{\mbox{\scriptsize bh}}k\delta(r)}{2\pi r}.
\end{equation}

Using this new one density-potential pair we find that
\begin{equation}
e^{\tilde{\lambda'}}(r,z)=\frac{k}{\tilde{U}(r,z) + k + M_{\mbox{\scriptsize bh}}k/\sqrt{r^2+z^2}},\label{rnela}
\end{equation}
and
\begin{equation}
\tilde{\epsilon}'(r)=-\frac{k \tilde{\Sigma}(r)}{(\tilde{U}(r,0) + k + M_{\mbox{\scriptsize bh}}k/r)^2}\label{rnepsi}
\end{equation}
plus a delta term at the center. Evidently, the case $\tilde{U}(r,z)=0$ reduces to the metric (\ref{rncyl}). It is interesting to note that the total masses of the disks are modified by the presence of the black hole. Now $\mathcal{M}$ depends on $M_{\mbox{\scriptsize bh}}$ in the following way:
\begin{equation}
\mathcal{M}=-2\pi a\int_0^{a}\frac{\Sigma(R)}{\frac{R}{a}U(R) + k+M_{\mbox{\scriptsize bh}}kaR^2}dR. \label{masabh}
\end{equation}

This new solution has a singularity at the origin. From the analysis performed in section \ref{sec:emt} it is clear that our solution is a non-linear superposition of an extremal RN black hole and a charged dust disk. Now, if we we look at the limit $r\rightarrow 0$ and $z\rightarrow 0$ we obtain
\begin{equation}
ds^2=-\left(\frac{r^2+z^2}{M_{\mbox{\scriptsize bh}}^2}\right)dt^2+\left(\frac{M_{\mbox{\scriptsize bh}}^2}{r^2+z^2}\right)(dr^2+dz^2+r^2d\varphi^2),
\end{equation}
or equivalently
\begin{equation}
ds^2=-\left(\frac{R^2}{M_{\mbox{\scriptsize bh}}^2}\right)dt^2+\left(\frac{M_{\mbox{\scriptsize bh}}^2}{R^2}\right)dR^2+M_{\mbox{\scriptsize bh}}^2d\Omega_2^2.
\end{equation}
This metric has the Robinson-Bertotti form $AdS_2\times S^2$ \cite{ROBI,BERT}, with anti--de Sitter ($AdS$) length $L=M_{\mbox{\scriptsize bh}}$, and coincides with the near-horizon limit of an extremal RN black hole, as expected.

\subsection{An explicit solution\label{exsol}}

In this subsection we present a fully integrated solution representing the superposition of the first member of inverted MM family with an extremal RN black hole. The obtention of other solutions can be done in a similar way and the physical properties are essentially the same. Therefore we will focus only in this model.

Perhaps, the simplest way to obtain a closed expression for the metric coefficient is to express the potential $U$ in terms of oblate spheroidal coordinates as was done in the previous sections. However, Kelvin transformation requires that we first express $U$ in cylindrical coordinates which can be done analyzing the patches $r<a$ and $r>a$ separately. A more convenient expression for the potential was given recently in \cite{EAR}, in which the author obtained it in a closed-form for all values of $r$ in cylindrical coordinates.

For the $m=1$ disk $U(r,z)$ is given by \cite{EAR}:
\begin{align}
U(r,z)=-\frac{3 M G}{8 a^3}\bigg[&2( 2a^2 -r^2 +2z^2)\sin^{-1}\left( \frac{f_1-f_2}{2 r} \right) \nonumber \\
&+\sqrt{2}a \sqrt{f_1 f_2 -f_3}-3\sqrt{2}|z|  \sqrt{f_1 f_2 +f_3}  \bigg],
\end{align}
where
\begin{align}
   f_1 &= \sqrt{z^2+(r +a)^2},  \nonumber  \\
   f_2 &= \sqrt{z^2+(r -a)^2},            \\
   f_3 &= a^2 -r^2 -z^2.        \nonumber
\end{align}
On the other hand, he wave that the corresponding surface density is
\begin{equation}
 \Sigma(r)  =   \begin{cases}  \frac{3M}{2\pi a^2} \sqrt{1-r^2/a^2}  & \textrm{for } r<a, \\
                             0                                                          & \textrm{for } r>a.
\end{cases}\end{equation}
Therefore, the Kelvin inverted potential-density pair reduces to
\begin{align}
\tilde{U}(r,z)=-\frac{3 M G}{8 a^2 \sqrt{r^2+z^2}}\bigg[&2\left( 2a^2 +\frac{a^4(2z^2-r^2)}{(r^2+z^2)^2}\right)\sin^{-1}\left( \frac{(f'_1-f'_2)(r^2+z^2)}{2 a^2 r} \right) \nonumber \\
&+\sqrt{2}a     \sqrt{f'_1 f'_2 -f'_3}-3\sqrt{2}\left|\frac{a^2 z}{r^2+z^2}\right|  \sqrt{f'_1 f'_2 +f'_3}  \bigg],\label{rnu}
\end{align}
where
\begin{align}
   f'_1 &= \sqrt{\frac{a^2(z^2+(r+a)^2)}{r^2+z^2}},  \nonumber  \\
   f'_2 &= \sqrt{\frac{a^2(z^2+(r-a)^2)}{r^2+z^2}},             \\
   f'_3 &= \frac{a^2(a^2 -r^2 -z^2)}{r^2+z^2},        \nonumber
\end{align}
and
\begin{equation}
\tilde{ \Sigma}(r)  =   \begin{cases} 0  & \textrm{for } r<a, \\
                             \frac{3aM}{2\pi r^3} \sqrt{1-a^2/r^2}                                                          & \textrm{for } r>a.\label{rns}
\end{cases}
\end{equation}

Plugging equations (\ref{rnu}) and (\ref{rns}) into (\ref{rnela}) and (\ref{rnepsi}), expressions for the metric coefficient and surface energy density of the disk can be found.

\begin{figure*}[!ht]
\begin{center}
\includegraphics[width=4.8in]{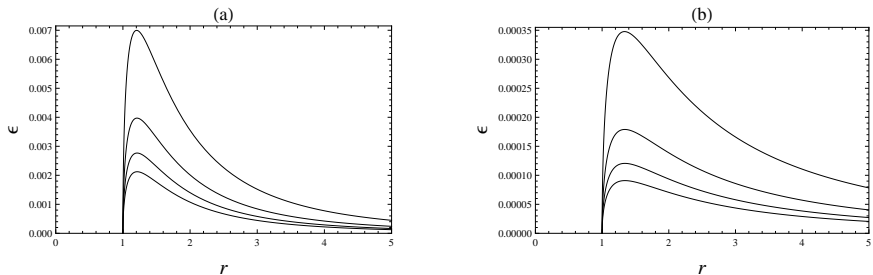}
\caption{\label{figure5}Dimensionless surface energy density ${\hat
\epsilon}$ as a function of ${\hat r}$ for the superposition of the first member of the inverted MM family with a black hole of mass (a) $M_{\mbox{\scriptsize bh}}=M$; (b) $M_{\mbox{\scriptsize bh}}=10M$. In each case, we plot ${\hat \epsilon}^{(m)} ({\hat r})$ for $0 \leq {\hat r} \leq 5$ with different values for the parameter ${\hat k}$. The uppest curve of each plot corresponds to ${\hat k} = -5$, and
then ${\hat k} = -10$, $-15$ and $-20$ for the lowest curve.}
\end{center}
\end{figure*}

\begin{figure*}[!ht]
\begin{center}
\includegraphics[width=7cm]{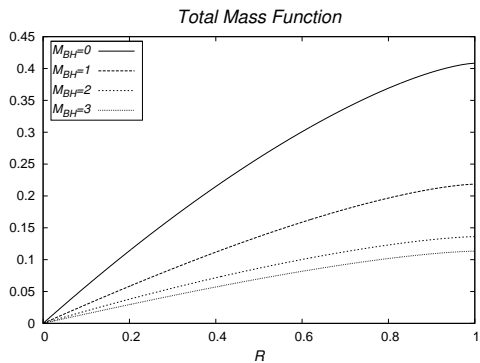}
\caption{In this plot, we present the mass function for the superposition of the first model with a black hole. Here we set $a=1$, $G=1$, $k=-5$ and $M=1$ while varying $M_{\mbox{\scriptsize bh}}$. The mass of the disk is reduced as the black hole mass is increased.}\label{fig:masabh}
\end{center}
\end{figure*}

In Figure \ref{figure5}, we plot the dimensionless surface energy density ${\hat\epsilon}$ as a function of $\hat{r}$ for the superposition of an extremal Reissner-Nordstr\"{o}m black hole with the first member of the inverted MM family. The physical behavior of the density profiles are pretty similar to the plots showed in Fig. \ref{fig:1}. In Fig. \ref{fig:masabh} we plotted the mass function for this particular model with different black hole masses and we found that the mass of the disk is reduced as the black hole mass is increased. This is precisely the expected behavior from Equation (\ref{masabh}).

\section{\label{sec:conc}Concluding Remarks}

We have presented the first superposition of a charged black hole with an annular disk made of extremal dust. To do this, we first obtained two infinite families of axially symmetric charged dust disks with a central hole, with well-behaved surface energy and charge densities. The disks were obtained by solving the electrovacuum Einstein-Maxwell equations system for conformastatic spacetimes, assuming a functional dependence between the metric function, the electric potential and an auxiliary function that was taken as a solution of the Laplace equation. Moreover, the solutions obtained here have a charge density that is equal, except maybe by a sign, to their mass density, in such a way that the electric and gravitational forces are in exact balance.

Then, we employed the well-known $Lord$ $Kelvin$ $Inversion$ $Method$ in order to obtain annular disks, applied to the MM and flat rings systems, which are models of finite extension. Here is worth to mention that although this method was already employed by Lemos and Letelier in a previous work \cite{LL2}, the corresponding metric functions could not be fully analytically computed in terms of elementary functions. The most important point to obtain the superposition of these disks with an extremal Reissner-Nordstr\"{o}m black hole was the realization that such geometry arises naturally within the inversion by considering an extra boundary term. The solutions obtained have positive and well-behaved energy densities, vanishing at the edge, and as the disks are made of dust, all the models are in complete agreement with all the energy conditions, a fact of particular relevance in the study of relativistic thin disks models. Indeed, as was mentioned at the introduction, many of the relativistic thin disks models that had been studied in the literature do not fully agrees with these conditions. Furthermore, although the structures obtained are extended to infinity, one one can put a clear cutoff due to the fast decay rate of the densities, and their masses are finite. This same features hold for the solutions with or without the central black hole.

We believe that the obtained models have
some remarkable properties; its
relative simplicity and the clear identification of the material source
associated to it not being the least of them. In particular, they may be suitable to study their generalizations to conformastatic models in the presence of magnetic fields as well as conformastationary models with electromagnetic fields, so we are now considering some research in this direction. Also the study of the dynamics of test fields over these backgrounds is under consideration.

\section*{Acknowledgments}

FDL-C and JFP want to thank Mexico's National Council of Science and Technology (CONACyT) for financial support.


\begin{thebibliography}{99}
\bibitem{comment} Here we have used the identity in cylindrical coordinates $\delta(\mathbf{x})=\delta(r)\delta(z)/(2\pi r)$ and interpreted the term $\delta(r)/(2\pi r)$ as an effective "surface density".
\end{thebibliography}
\end{document}